\documentclass[aps,prr,a4paper,superscriptaddress,twocolumn,showpacs,amsmath,amssymb,longbibliography]{revtex4-2}

\usepackage{graphicx,epsfig}
\usepackage[usenames]{color}

\usepackage[english]{babel}

\usepackage{dsfont}

\usepackage{csquotes}
\MakeOuterQuote{"}

\allowdisplaybreaks

\begin{document}

\newcommand{\E}{\mathcal{E}}
\newcommand{\G}{\mathcal{G}}
\newcommand{\Lag}{\mathcal{L}}
\newcommand{\M}{\mathcal{M}}
\newcommand{\N}{\mathcal{N}}
\newcommand{\U}{\mathcal{U}}
\newcommand{\R}{\mathcal{R}}
\newcommand{\F}{\mathcal{F}}
\newcommand{\V}{\mathcal{V}}
\newcommand{\C}{\mathcal{C}}
\newcommand{\I}{\mathcal{I}}
\newcommand{\s}{\sigma}
\newcommand{\up}{\uparrow}
\newcommand{\dw}{\downarrow}
\newcommand{\h}{\hat{\mathcal{H}}}
\newcommand{\himp}{\hat{h}}
\newcommand{\g}{\mathcal{G}^{-1}_0}
\newcommand{\D}{\mathcal{D}}
\newcommand{\A}{\mathcal{A}}
\newcommand{\projs}{\hat{\mathcal{S}}_d}
\newcommand{\proj}{\hat{\mathcal{P}}_d}
\newcommand{\K}{\textbf{k}}
\newcommand{\Q}{\textbf{q}}
\newcommand{\T}{\tau_{\ast}}
\newcommand{\io}{i\omega_n}
\newcommand{\eps}{\varepsilon}
\newcommand{\+}{\dag}
\newcommand{\su}{\uparrow}
\newcommand{\giu}{\downarrow}
\newcommand{\0}[1]{\textbf{#1}}
\newcommand{\ca}{c^{\phantom{\dagger}}}
\newcommand{\cc}{c^\dagger}

\newcommand{\pa}{{p}^{\phantom{\dagger}}}
\newcommand{\pc}{{p}^\dagger}

\newcommand{\fa}{f^{\phantom{\dagger}}}
\newcommand{\fc}{f^\dagger}
\newcommand{\aaa}{a^{\phantom{\dagger}}}
\newcommand{\aac}{a^\dagger}
\newcommand{\bba}{b^{\phantom{\dagger}}}
\newcommand{\bbc}{b^\dagger}
\newcommand{\da}{{d}^{\phantom{\dagger}}}
\newcommand{\dc}{{d}^\dagger}
\newcommand{\ha}{h^{\phantom{\dagger}}}
\newcommand{\hc}{h^\dagger}
\newcommand{\be}{\begin{equation}}
\newcommand{\ee}{\end{equation}}
\newcommand{\bea}{\begin{eqnarray}}
\newcommand{\eea}{\end{eqnarray}}
\newcommand{\ba}{\begin{eqnarray*}}
\newcommand{\ea}{\end{eqnarray*}}
\newcommand{\dagga}{{\phantom{\dagger}}}
\newcommand{\bR}{\mathbf{R}}
\newcommand{\bQ}{\mathbf{Q}}
\newcommand{\bq}{\mathbf{q}}
\newcommand{\bqp}{\mathbf{q'}}
\newcommand{\bk}{\mathbf{k}}
\newcommand{\bh}{\mathbf{h}}
\newcommand{\bkp}{\mathbf{k'}}
\newcommand{\bp}{\mathbf{p}}
\newcommand{\bL}{\mathbf{L}}
\newcommand{\bRp}{\mathbf{R'}}
\newcommand{\bx}{\mathbf{x}}
\newcommand{\by}{\mathbf{y}}
\newcommand{\bz}{\mathbf{z}}
\newcommand{\br}{\mathbf{r}}
\newcommand{\Ima}{{\Im m}}
\newcommand{\Rea}{{\Re e}}
\newcommand{\Pj}[2]{|#1\rangle\langle #2|}
\newcommand{\ket}[1]{\vert#1\rangle}
\newcommand{\bra}[1]{\langle#1\vert}
\newcommand{\setof}[1]{\left\{#1\right\}}
\newcommand{\fract}[2]{\frac{\displaystyle #1}{\displaystyle #2}}
\newcommand{\Av}[2]{\langle #1|\,#2\,|#1\rangle}
\newcommand{\av}[1]{\langle #1 \rangle}
\newcommand{\Mel}[3]{\langle #1|#2\,|#3\rangle}
\newcommand{\Avs}[1]{\langle \,#1\,\rangle_0}
\newcommand{\eqn}[1]{(\ref{#1})}
\newcommand{\Tr}{\mathrm{Tr}}

\newcommand{\Vb}{\bar{\mathcal{V}}}
\newcommand{\Vd}{\Delta\mathcal{V}}
\def\P{P_{02}}
\newcommand{\Pb}{\bar{P}_{02}}
\newcommand{\Pd}{\Delta P_{02}}
\def\t{\theta_{02}}
\newcommand{\tb}{\bar{\theta}_{02}}
\newcommand{\td}{\Delta \theta_{02}}
\newcommand{\Rb}{\bar{R}}
\newcommand{\Rd}{\Delta R}
\newcommand{\ocrev}[1]{{\color{cyan}{#1}}}
\newcommand{\occom}[1]{{\color{red}{#1}}}

\newcommand{\Dp}{\hat{\Delta}_{\text{p}}}
\newcommand{\Dh}{\hat{\Delta}_{\text{h}}}
\newcommand{\Deltap}{[\hat{\Delta}_{\text{p}}]}
\newcommand{\Deltah}{[\hat{\Delta}_{\text{h}}]}

\title{Operatorial formulation of the ghost rotationally-invariant slave-Boson theory}


\author{Nicola Lanat\`a}
\altaffiliation{Corresponding author: lanata@phys.au.dk}
\affiliation{Department of Physics and Astronomy, Aarhus University, 8000, Aarhus C, Denmark}
\affiliation{Nordita, KTH Royal Institute of Technology and Stockholm University, Hannes Alfv\'ens v\"ag 12, SE-106 91 Stockholm, Sweden}

\date{\today}

\begin{abstract}

%

We propose an operatorially-exact formalism for describing the equilibrium and quantum-dynamical properties of many-electron systems interacting locally on a lattice, called “ghost rotationally-invariant slave-Boson theory” (g-RISB).
We demonstrate that our theoretical framework reduces to the recently-developed ghost Gutzwiller approximation (g-GA) at the mean-field level.
Furthermore, we introduce the time-dependent mean-field g-RISB action, generalizing the time-dependent GA theory.
Since the g-RISB is based on exact reformulation of the many-body problem, it may pave the way to the development of practical implementations for adding systematically quantum-fluctuation corrections towards the exact solution, in arbitrary dimension.

\end{abstract}

\maketitle

\section{Introduction}

The idea of utilizing subsidiary degrees of freedom for modeling the strong interactions in many-electron systems has a long history in condensed-matter physics~\cite{subsidiary-He}, and it is nowadays imbued within numerous theoretical frameworks, such as tensor-networks methods~\cite{subsidiary-MPS,subsidiary-TN}, neural-network quantum states~\cite{subsidiary-NN}, slave-boson methods~\cite{SB1,SB2,SB3,SB4,SB5,SB6,risb-georges,Lanata2016} and the recently-developed g-GA~\cite{Ghost-GA,ALM_g-GA,Guerci}.
One of such frameworks, which proved to be particularly useful within the context of ab-initio real-material calculations, is the so-called rotationally-invariaint slave-Boson theory (RISB)~\cite{risb-georges,Lanata2016}.
So far, real-material applications of the RISB method have been mostly limited to the mean-field level~\cite{Ho,Fang,Lanata2016,npj-lanata,Slave-bosons-materials,Ho,PhysRevLett.110.096401,PhysRevLett.104.047002,PhysRevLett.108.036406,PhysRevLett.105.096401,FeSe-FeTe, Our-Ce}, ---which is equivalent to the multi-orbital Gutzwiller approximation (GA)~\cite{Gutzwiller3,equivalence_GA-SB,lanata-barone-fabrizio}, and can be formulated as a quantum-embedding method~\cite{quantum-embedding-review,Kotliar-Science}, see Ref.~\cite{Our-PRX}.
A key reason at the basis of the success of the RISB mean-field theory is that it is much less computationally demanding compared to other methods, such as dynamical mean-field theory (DMFT)~\cite{DMFT,dmft_book}.
On the other hand, the accuracy of the RISB mean-field approximation is not always sufficient; in fact, it can be even incorrect by orders of magnitude in some parameters regimes~\cite{ALM_g-GA}.
Therefore, systematic methods for improving its accuracy are desirable.

A possible way of overcoming the limitations of the RISB mean-field theory is the g-GA~\cite{Ghost-GA,ALM_g-GA,Guerci}, which is based on the idea of extending the variational space by introducing auxiliary Fermionic degrees of freedom.
The key advantage of the g-GA is that it proved to have accuracy comparable with DMFT, at a much lower computational cost.
However, from the accuracy standpoint, this method still requires to make approximations that become exact only in the limit of infinite dimension.
Therefore, its current formulation (that is based on extending the Gutzwiller wavefunction, rather than a slave-boson perspective) does not provide tools for including systematic corrections towards the exact solution in low-dimensional systems (such as those necessary for capturing the non-local correlation effects).
Another way of improving the accuracy of the RISB mean-field solution is to take into account perturbatively the quantum-fluctuation corrections. In fact, a rigorous operatorial formulation of the RISB theory (reducing to GA at the mean-field level) exists~\cite{Lanata2016}, and previous work showed that taking into account the
quantum fluctuations allows one to capture effective interactions (mediated by the slave bosons) among the quasiparticles defined at the saddle point level~\cite{SB-fluct-1}.
However, since the mean-field starting point can be qualitatively inaccurate in the strongly-correlated regime~\cite{ALM_g-GA}, 
including quantum-fluctuation corrections perturbatively may not be sufficient, in general.

Here we combine the two formalisms mentioned above (the g-GA and the RISB).
Specifically, we generalize the mathematical constructions of Refs.~\cite{risb-georges} and \cite{Lanata2016} to design a 
slave-boson theory (the g-RISB) that: (i) is an exact operatorial  representation of the many-body problem and (ii) reduces to the g-GA at the mean-field level.
Since the g-RISB mean-field solution (i.e., the g-GA) 
describes the electronic structure in terms of emergent Bloch excitations~\cite{Ghost-GA}, and such description
proved to have accuracy comparable with DMFT in all parameters regimes~\cite{Ghost-GA,ALM_g-GA} (including the Mott phase),
our formalism may pave the way to implementations able to take into account perturbatively the residual effective interactions between such generalized emergent states, allowing us to perform high-precision calculations of strongly-correlated electron systems, in arbitrary dimension.

\section{The model}

We consider a generic multi-orbital Fermionic Hamiltonian represented as follows:
\begin{align}
\hat{H} =\! \sum_{\bR\bRp ij} 
\sum_{\alpha=1}^{{\nu_i}} \!
\sum_{\beta=1}^{{\nu_j}} \!
{t}_{\bR\bRp,ij}^{\alpha\beta}\,
\cc_{\bR i\alpha}\ca_{\bRp j\beta} \!+\!
\sum_{\bR}\!
\hat{H}_{\bR i}^{\text{loc}}[\cc_{\bR i\alpha},\ca_{\bR i\alpha}]
,
\label{hubb}
\end{align}
where $\bR$ indicates the unit-cell (which is assumed to be repeated periodically), 
while $i$ represents different groups of degrees of freedom within each unit cell (such as orbital shells) and $\alpha$ labels the spin and orbital degrees of freedom for each $i$.
Each local subsystem $\mathcal{H}_{\bR i}$ generated by the Fermionic modes $\{\cc_{\bR i\alpha}|\,\alpha=1,..,\nu_i\}$ is spanned by the following $2^{\nu_i}$ 
Fock states:
\be
\ket{\Gamma,{\bR i}}=[\cc_{\bR i1}]^{q_1(\Gamma)} ...
[\cc_{\bR i q_{\nu_i}}]^{q_{\nu_i}(\Gamma)}\,\ket{0}
\label{ket-GammaRi}
\,,
\ee
where $\Gamma\in\{0,..,2^{\nu_i}-1\}$ and $q_l(\Gamma)$ is the $l$-th digit of $\Gamma$ in binary representation.
With this notation, the full physical space of the system can be represented as follows:
\be
\mathcal{H}=\bigotimes_{\bR i}\mathcal{H}_{\bR i}\,,
\ee
and the local terms of Eq.~\eqref{hubb}, that here we assume to conserve the number of electrons, can be rewritten as:
\begin{align}
\hat{H}_{\bR i}^{\text{loc}}
&=\sum_{\Gamma,\Gamma'=0}^{2^{\nu_i}-1}
[{H}_{ i}^{\text{loc}}]_{\Gamma\Gamma'}
\,\ket{\Gamma, \bR i}\bra{\Gamma', \bR i}
\\
[{H}_{ i}^{\text{loc}}]_{\Gamma\Gamma'}
&=\langle\Gamma, \bR i |
\hat{H}_{\bR i}^{\text{loc}}
| \Gamma', \bR i\rangle
\,,
\label{hloc-matrix}
\end{align}
where the matrix ${H}_{i}^{\text{loc}}$ does not depend on the unit-cell label $\bR$ because of the translational invariance of $\hat{H}$.

From now on, with no loss of generality, we assume that ${t}_{\bR\bRp,ii}^{\alpha\beta}=0$ (as in Refs.~\cite{Our-PRX} and \cite{Lanata2016}), i.e., we include all on-site single-particle terms within the definition of $\hat{H}_{\bR i}^{\text{loc}}$.

In the next section we are going to show that Eq.~\eqref{hubb} can be equivalently reformulated utilizing auxiliary Bosons and Fermions.

\section{Operatorial $\text{g}$-RISB formulation of the many-body problem}

Let us consider the Fock space $\underline{\mathcal{F}}$ generated by a set of auxiliary Fermionic modes:
\be
\{\fc_{\bR ia}|\,a=1,..,\mathcal{B}_i\nu_i\}\,,
\ee
where $\mathcal{B}_i\geq 1$ is a given integer,
and a set of Bosonic modes:
\begin{align}
\big\{
\Phi_{\bR i \Gamma n}\,|\,n\in S^\Gamma_{i}\,\forall\,
\Gamma\in\{0,..,2^{\nu_i}-1\}
\big\} \,,
\end{align}
where:
\begin{align}
S^\Gamma_{i}&=
\big\{
n\in\{0,..,2^{\mathcal{B}_i\nu_i}-1\}
\,|\,
N(n)-N(\Gamma)=\mathcal{M}_i
\label{SGammai}
\big\}
\\
N(\Gamma)&=\sum_{\alpha=1}^{\nu_i}q_\alpha(\Gamma)
\\
N(n)&=\sum_{a=1}^{\mathcal{B}\nu_i}q_a(n)
\label{Nofn}
\end{align}
and:
\be
\mathcal{M}_i=\frac{\nu_i}{2}(\mathcal{B}_i-1)
\,.
\label{half-filling-EH}
\ee
Throughout the rest of this paper we will
assume that $\mathcal{B}_i=\mathcal{B}$ is independent of $i$.
The reason underlying Eq.~\eqref{half-filling-EH} is that, as we are going to show later, the resulting theory reduces to the g-GA at the mean-field level.
However, we note that different choices $\mathcal{B}$ and $\mathcal{M}_i$ would be also possible, in principle.
We will also adopt the convention that $\Phi_{\bR i \Gamma n}=0$ $\forall\,n\notin S^\Gamma_{i}$.

The mathematical structure defined above reduces to the standard RISB~\cite{risb-georges} for $\mathcal{B}=1$, while it features additional auxiliary Fermionic and Bosonic modes for $\mathcal{B}>1$.
As we are going to show below, it is possible to construct an exact alternative reformulation of the many-body problem $\forall\,\mathcal{B}\geq 1$, in such a way that the g-GA is recovered at the mean-field level.
The resulting generalized framework will be denoted g-RISB.

\subsection{The physical subspace}

We consider the following subspace of the Fock space introduced above:
\be
\underline{\mathcal{H}}^{\mathcal{B}}=\bigotimes_{\bR i}
\underline{\mathcal{H}}_{\bR i}^{\mathcal{B}}\,,
\ee
where $\underline{\mathcal{H}}_{\bR i}^{\mathcal{B}}$ is spanned the following states:
\begin{align}
    \ket{\underline{\Gamma},\bR i}&=
    D_{i\Gamma}^{-\frac{1}{2}}
    \sum_{n\in S^\Gamma_{i}}
    \Phi^\dagger_{\bR i \Gamma n}\,\ket{n,\bR i}
    \label{ket-GammanRi}
    \\
    \ket{n,\bR i}&=
    [\fc_{\bR i1}]^{q_1(n)} ...
    [\fc_{\bR i q_{\mathcal{B}\nu_i}}]^{q_{\mathcal{B}\nu_i}(n)}\,\ket{0}
    \,,
    \label{ket-nRi}
\end{align}
where:
\be
D_{i\Gamma}=
\frac{(\mathcal{B}\nu_i)!}{(N(\Gamma)+\mathcal{M}_i)!\, (\mathcal{B}\nu_i-N(\Gamma)-\mathcal{M}_i)!}
\label{normalizationD}
\ee
is a normalization factor
constructed in such a way that:
\be
\langle\underline{\Gamma},\bR i | \underline{\Gamma}',\bR' i'\rangle
=\delta_{\bR\bR'}\delta_{ii'}\delta_{\Gamma\Gamma'}
\,.
\ee

All states in $\underline{\mathcal{H}}^{\mathcal{B}}$
satisfy (and are fully characterized by) the so-called Gutzwiller constraints:
\begin{align}
    \underline{K}^0_{\bR i}&=
    \sum_{\Gamma=0}^{2^{\nu_i}-1} \sum_{n=0}^{2^{\mathcal{B}\nu_i}-1}
    \Phi^\dagger_{\bR i \Gamma n}
    \Phi^\dagga_{\bR i \Gamma n} \!-\! 1 =0
    \label{gc1}
    \\
    \underline{K}_{\bR i}^{ab}&=
    \sum_{\Gamma=0}^{2^{\nu_i}-1} \sum_{n,m=0}^{2^{\mathcal{B}\nu_i}-1}
    \big[\tilde{F}_{ia}^\dagger \tilde{F}_{ib}^\dagga\big]_{mn}
    \Phi^\dagger_{\bR i \Gamma n}
    \Phi^\dagga_{\bR i \Gamma m}
    \nonumber\\&\qquad\qquad\qquad\qquad\qquad
    - \fc_{\bR ia} \fa_{\bR ib} =0
    \,,
    \label{gc2}
\end{align}
where $a,b\in\{1,..,\mathcal{B}\nu_i\}$ and
we introduced the following $\mathcal{B}\nu_i\times \mathcal{B}\nu_i$ matrix representations of the auxiliary Fermionic operators:
\begin{align}
[\tilde{F}^\dagger_{ia}]_{nn'}= 
\langle {n},\bR i | 
\fc_{\bR i a}
| {n}',\bR i\rangle
\,.
\label{reprtildeF}
\end{align}
Note that the entries of the matrix representations $\tilde{F}^\dagger_{ia}$ do not depend on the unit-cell label $\bR$, because it appears both in $\fc_{\bR i a}$ and in the definition of the Fock states $\ket{{n},\bR i}$, see Eq.~\eqref{ket-nRi}.

For completeness, let us show explicitly that the states defined in Eq.~\eqref{ket-GammanRi}, which generate the physical subspace, satisfy the Gutzwiller constraints.
The Eq.~\eqref{gc1} follows immediately from the observation that, by definition, all physical states have a single Boson.
The proof of Eq.~\eqref{gc2} is the following:
\begin{widetext}
\begin{align}
    &\sum_{\Gamma'=0}^{2^{\nu_i}-1} \sum_{n',m'=0}^{2^{\mathcal{B}\nu_i}-1}
    \big[\tilde{F}_{ia}^\dagger \tilde{F}_{ib}^\dagga\big]_{m'n'}
    \Phi^\dagger_{\bR i \Gamma' n'}
    \Phi^\dagga_{\bR i \Gamma' m'}\,\ket{\underline{\Gamma},\bR i}
    \nonumber\\
    &\qquad= \sum_{\Gamma'=0}^{2^{\nu_i}-1} \sum_{n',m'=0}^{2^{\mathcal{B}\nu_i}-1}
    \big[\tilde{F}_{ia}^\dagger \tilde{F}_{ib}^\dagga\big]_{m'n'}
    \Phi^\dagger_{\bR i \Gamma' n'}
    \Phi^\dagga_{\bR i \Gamma' m'}\,
    D_{i\Gamma}^{-\frac{1}{2}}
    \sum_{n\in S^\Gamma_{i}}
    \Phi^\dagger_{\bR i \Gamma n}\,\ket{n,\bR i}
    \nonumber\\
    &\qquad= D_{i\Gamma}^{-\frac{1}{2}}
    \sum_{\Gamma'=0}^{2^{\nu_i}-1} \sum_{n',m'=0}^{2^{\mathcal{B}\nu_i}-1}
    \big[\tilde{F}_{ia}^\dagger \tilde{F}_{ib}^\dagga\big]_{m'n'}
    \Phi^\dagger_{\bR i \Gamma' n'}
    \sum_{n\in S^\Gamma_{i}}
    \delta_{\Gamma\Gamma'}\delta_{nm'}
    \,\ket{n,\bR i}
    \nonumber\\
    &\qquad= D_{i\Gamma}^{-\frac{1}{2}}
    \sum_{n,n'\in S^\Gamma_{i}}
    \big[\tilde{F}_{ia}^\dagger \tilde{F}_{ib}^\dagga\big]_{nn'}
    \Phi^\dagger_{\bR i \Gamma' n'}
    \, \ket{n,\bR i}
    \nonumber\\
    &\qquad= D_{i\Gamma}^{-\frac{1}{2}}
    \sum_{n'\in S^\Gamma_{i}}
    \Phi^\dagger_{\bR i \Gamma' n'}
    \sum_{n\in S^\Gamma_{i}}
    \, \ket{n,\bR i}
    \langle n,\bR i | \fc_{\bR ia}\fa_{\bR ib} | n',\bR i \rangle
    \nonumber\\
    &\qquad= D_{i\Gamma}^{-\frac{1}{2}}
    \sum_{n'\in S^\Gamma_{i}}
    \Phi^\dagger_{\bR i \Gamma' n'}
    \fc_{\bR ia}\fa_{\bR ib}\ket{n',\bR i}=
    \fc_{\bR ia}\fa_{\bR ib}
    \,\ket{\underline{\Gamma},\bR i}
    \,,
\end{align}
\end{widetext}
where we used that, because of Eq.~\eqref{reprtildeF},
$
\big[\tilde{F}_{ia}^\dagger \tilde{F}_{ib}^\dagga\big]_{nn'}
=\langle n,\bR i | \fc_{\bR ia}\fa_{\bR ib} | n',\bR i \rangle
$.


\subsection{Representation of local operators}

Let us consider the following operators in $\underline{\mathcal{H}}^{\mathcal{B}}$:
\begin{align}
\underline{\hat{H}}_{\bR i}^{\text{loc}}
&=\sum_{\Gamma,\Gamma'=0}^{2^{\nu_i}-1}
[{H}_{ i}^{\text{loc}}]_{\Gamma\Gamma'}
\,
\sum_{n\in S^\Gamma_{i}} 
\Phi^\dagger_{\bR i \Gamma n}
\Phi^\dagga_{\bR i \Gamma' n}
\\
[{H}_{ i}^{\text{loc}}]_{\Gamma\Gamma'}
&=\langle\Gamma, \bR i |
\hat{H}_{\bR i}^{\text{loc}}
| \Gamma', \bR i\rangle
\,,
\end{align}
where $S^\Gamma_{i}$ was defined in Eq.~\eqref{SGammai} and the
the matrix ${H}_{i}^{\text{loc}}$ was previously defined in Eq.~\eqref{hloc-matrix}.

It can be readily verified by inspection that:
\begin{align}
\langle\underline{\Gamma},\bR i | 
\underline{\hat{H}}_{\bR i}^{\text{loc}}
| \underline{\Gamma}',\bR i\rangle
&=
\langle{\Gamma},\bR i | 
{\hat{H}}_{\bR i}^{\text{loc}}
|{\Gamma}',\bR i\rangle
\nonumber
\\
&= [{H}_{ i}^{\text{loc}}]_{\Gamma\Gamma'}
\;\,
\forall\,\Gamma,\Gamma'
\,.
\end{align}
In fact:
\begin{widetext}

\begin{align}
\langle\underline{\Gamma},\bR i | 
\underline{\hat{H}}_{\bR i}^{\text{loc}}
| \underline{\Gamma}',\bR i\rangle
&=
\sum_{\Gamma_1,\Gamma_2=0}^{2^{\nu_i}-1}
[{H}_{ i}^{\text{loc}}]_{\Gamma_1\Gamma_2}
\langle\underline{\Gamma},\bR i | 
\sum_{n=0}^{2^{\mathcal{B}\nu_i}-1}
\Phi^\dagger_{\bR i \Gamma_1 n}
\Phi^\dagga_{\bR i \Gamma_2 n}
| \underline{\Gamma}',\bR i\rangle
\nonumber\\
&=
(D_{i\Gamma}D_{i\Gamma'})^{-\frac{1}{2}}
\sum_{\Gamma_1,\Gamma_2=0}^{2^{\nu_i}-1}
[{H}_{ i}^{\text{loc}}]_{\Gamma_1\Gamma_2}
\sum_{n,m,m'\in S^\Gamma_{i}}
\langle{m},\bR i | 
\Phi^\dagga_{\bR i \Gamma m}\,
\Phi^\dagger_{\bR i \Gamma_1 n}
\Phi^\dagga_{\bR i \Gamma_2 n}\,
\Phi^\dagger_{\bR i \Gamma' m'}
| {m'},\bR i\rangle
\nonumber\\
&=
D_{i\Gamma}^{-1}
\sum_{\Gamma_1,\Gamma_2=0}^{2^{\nu_i}-1}
[{H}_{ i}^{\text{loc}}]_{\Gamma_1\Gamma_2}
\sum_{n,m,m'\in S^\Gamma_{i}}
\delta_{\Gamma\Gamma_1}\delta_{nm}\delta_{\Gamma_2\Gamma'}\delta_{nm'}
=[{H}_{ i}^{\text{loc}}]_{\Gamma\Gamma'}
\,,
\end{align}
\end{widetext}
where we used that $N(\Gamma_1)=N(\Gamma_2)$ (which is true because $\hat{H}$ was assumed to conserve the number of electrons).

This shows that $\underline{\hat{H}}_{\bR i}^{\text{loc}}$ is an exact
equivalent representation of ${\hat{H}}_{\bR i}^{\text{loc}}$
within $\underline{\mathcal{H}}^{\mathcal{B}}$.

\subsection{Representation of one-body non-local operators}

Let us show that it is possible to construct a set of operators
$\{\hat{\mathcal{R}}_{\bR i a\alpha}\,|\,\alpha\in\{1,..,\nu_i\},a\in\{1,..,\mathcal{B}\nu_i\}\}$
in such a way that:
\be
\underline{c}^\dagger_{\bR i \alpha}=
\sum_{a=1}^{\mathcal{B}\nu_i} \hat{\mathcal{R}}_{\bR i a\alpha}
\fc_{\bR i a}
\label{equivalence-general}
\ee
satisfy the following equality:
\begin{align}
\langle\underline{\Gamma},\bR i | 
\underline{c}^\dagger_{\bR i \alpha}
| \underline{\Gamma}',\bR i\rangle
&=
\langle{\Gamma},\bR i | 
\cc_{\bR i \alpha}
|{\Gamma}',\bR i\rangle
\nonumber\\
=[F^\dagger_{i\alpha}]_{\Gamma \Gamma'}
\;\,
\forall\,\Gamma,\Gamma'
\,,
\label{Rcondition}
\end{align}
where we introduced the $\nu_i\times\nu_i$ matrix representations ${F}^\dagger_{i\alpha}$ of the physical Fermionic operators $\cc_{\bR i \alpha}$
(that, as the representations of Eq.~\eqref{reprtildeF}, do not depend on the unit-cell label $\bR$).

The condition [Eq.~\eqref{Rcondition}] can be realized using the following operators:
\begin{align}
\hat{\mathcal{R}}_{\bR i a\alpha}\!&=\!\!\!\!\!
\sum_{\Gamma\!_1 \!,\Gamma\!_2=0}^{2^{\nu_i}\!-\!1}
\sum_{n_1 \!,n_2=0}^{2^{\mathcal{B}\nu_i}\!-\!1}\!\!\!\!\!
C_{i,\Gamma\!_1 \! \Gamma\!_2}^{-\frac{1}{2}}
[F^\dagger_{i\alpha}]_{\Gamma\!_1 \! \Gamma\!_2}
[\tilde{F}^\dagger_{ia}]_{n_1 \! n_2}
\Phi^\dagger_{\bR i \Gamma\!_1  n_1}
\Phi^\dagga_{\bR i \Gamma\!_2  n_2}
\label{quadratic-R}
\\
C_{i,\Gamma_1\Gamma_2}\!&=
\left(N(\Gamma_1)+\mathcal{M}_i\right)
\big(\nu_i+\mathcal{M}_i-N(\Gamma\!_2)\big)
\,.
\label{quadratic-C}
\end{align}
In fact:
\begin{widetext}
\begin{align}
\langle\underline{\Gamma},\bR i | 
\underline{c}^\dagger_{\bR i \alpha}
| \underline{\Gamma}',\bR i\rangle
&=
\langle\underline{\Gamma},\bR i | 
\sum_{a=1}^{\mathcal{B}\nu_i} 
\hat{\mathcal{R}}_{\bR i a\alpha}
\fc_{\bR i a}| \underline{\Gamma}',\bR i\rangle
\nonumber\\&=
(D_{i\Gamma}D_{i\Gamma'})^{-\frac{1}{2}}
\sum_{a=1}^{\mathcal{B}\nu_i} 
\sum_{m\in S^\Gamma_{i}}
\sum_{m'\in S^{\Gamma'}_{i}}
\langle{m},\bR i | 
\Phi^\dagga_{\bR i \Gamma m}\,
\hat{\mathcal{R}}_{\bR i a\alpha}
\Phi^\dagger_{\bR i \Gamma' m'}
\fc_{\bR i a}
| {m'},\bR i\rangle
\nonumber\\ &=
(D_{i\Gamma}D_{i\Gamma'})^{-\frac{1}{2}}
C_{i,\Gamma \Gamma'}^{-\frac{1}{2}}\,
[F^\dagger_{i\alpha}]_{\Gamma \Gamma'}
\sum_{a=1}^{\mathcal{B}\nu_i} 
\sum_{m\in S^\Gamma_{i}}
\sum_{m'\in S^{\Gamma'}_{i}}
[\tilde{F}^\dagger_{ia}]_{m m'}
\langle{m},\bR i | 
\fc_{\bR i a}
| {m'},\bR i\rangle
\nonumber\\ &=
[F^\dagger_{i\alpha}]_{\Gamma \Gamma'}\,
(D_{i\Gamma}D_{i\Gamma'})^{-\frac{1}{2}}
C_{i,\Gamma  \Gamma'}^{-\frac{1}{2}}
\sum_{a=1}^{\mathcal{B}\nu_i} 
\sum_{m\in S^\Gamma_{i}}
[\tilde{F}^\dagger_{ia}\tilde{F}^\dagga_{ia}]_{m m}
\nonumber\\ &=
[F^\dagger_{i\alpha}]_{\Gamma \Gamma'}\,
(D_{i\Gamma}D_{i\Gamma'})^{-\frac{1}{2}}
C_{i,\Gamma \Gamma'}^{-\frac{1}{2}}
\sum_{m\in S^\Gamma_{i}}
N(m)
\nonumber\\ &=
[F^\dagger_{i\alpha}]_{\Gamma \Gamma'}\,
(D_{i\Gamma}D_{i\Gamma'})^{-\frac{1}{2}}
C_{i,\Gamma \Gamma'}^{-\frac{1}{2}}
\sum_{m\in S^\Gamma_{i}}
(N(\Gamma)+\mathcal{M}_i)
\nonumber\\ &=
[F^\dagger_{i\alpha}]_{\Gamma \Gamma'}\,
(D_{i\Gamma}D_{i\Gamma'})^{-\frac{1}{2}}
C_{i,\Gamma \Gamma'}^{-\frac{1}{2}}
D_{i\Gamma}
(N(\Gamma)+\mathcal{M}_i)
= [F^\dagger_{i\alpha}]_{\Gamma \Gamma'}
\,,
\end{align}
\end{widetext}
where the last step can be readily verified using the definitions in Eqs.~\eqref{half-filling-EH}, \eqref{normalizationD} and \eqref{quadratic-C}.

Note that Eq.~\eqref{quadratic-R} reduces to the RISB expression previously derived in Ref.~\cite{risb-georges} for the special case $\mathcal{B}=1$.

As we are going to show below following the procedure of Ref.~\cite{Lanata2016}, the equation for the renormalization operators $\hat{\mathcal{R}}_{\bR i a\alpha}$ can be modified in such a way that:
(i) the mapping between Eq.~\eqref{hubb} and Eq.~\eqref{hubb-risb} remains exact $\forall\,\mathcal{B}$ and, at the same time,
(ii) the resulting theory reduces to the recently-developed g-GA  $\forall\,\mathcal{B}>1$.
%
This is accomplished by the following expression:
\begin{widetext}
\begin{align}
\hat{\mathcal{R}}_{\bR i a\alpha}=
\sum_{\Gamma,\Gamma'=0}^{2^{\nu_i}-1}
\sum_{n,n'=0}^{2^{\mathcal{B}\nu_i}-1}
\sum_{b=1}^{\mathcal{B}\nu_i}
C_{i,\Gamma\Gamma'}^{-\frac{1}{2}}
[F^\dagger_{i\alpha}]_{\Gamma\Gamma'}
[\tilde{F}^\dagger_{ib}]_{nn'}
:
\Phi^\dagger_{\bR i \Gamma n}
&
\left[
\hat{\mathds{1}}+
\big(
C_{i,\Gamma\Gamma'}^{\frac{1}{2}}-1
\big)
\sum_{\Gamma''=0}^{2^{\nu_i}-1}
\sum_{n''=0}^{2^{\mathcal{B}\nu_i}-1}
\Phi^\dagger_{\bR i \Gamma'' n''}
\Phi^\dagga_{\bR i \Gamma'' n''}
\right]
\nonumber
\\&\quad
\left[
[\hat{\mathds{1}}-\hat{\Delta}_p]^{\left[-\frac{1}{2}\right]}
\bullet
[\hat{\mathds{1}}-\hat{\Delta}_h]^{\left[-\frac{1}{2}\right]}
\right]_{ba}
\Phi^\dagga_{\bR i \Gamma' n'}:
\,,
\label{R-operatorial}
\end{align}
\end{widetext}
where:
\begin{align}
\Deltap_{\bR i ab} &=
\sum_{\Gamma=0}^{2^{\nu_i}-1} \sum_{n,m=0}^{2^{\mathcal{B}\nu_i}-1} 
[\tilde{F}^\dagger_{ia}\tilde{F}^\dagga_{ib}]_{mn}\,
\Phi^\dagger_{\bR i \Gamma n}\Phi^\dagga_{\bR i \Gamma m}
\\
\Deltah_{\bR i ab} &=
\sum_{\Gamma=0}^{2^{\nu_i}-1} \sum_{n,m=0}^{2^{\mathcal{B}\nu_i}-1}  
[\tilde{F}^\dagga_{ib}\tilde{F}^\dagger_{ia}]_{mn}\,
\Phi^\dagger_{\bR i \Gamma n}\Phi^\dagga_{\bR i \Gamma m}
\,,
\end{align}
$\hat{\mathds{1}}$ is the identity operator,
the symbols $::$ indicate the normal ordering and we introduced the following opearators:
\bea
\left[\hat{\mathds{1}}-\Dp\right]^{[-\frac{1}{2}]} &\equiv&
\sum_{r=0}^\infty (-1)^r \binom{\frac{1}{2}}{r}\,\Deltap^{[r]}
\label{Dp-op}
\\
\left[\hat{\mathds{1}}-\Dh\right]^{[-\frac{1}{2}]} &\equiv&
\sum_{r=0}^\infty (-1)^r \binom{\frac{1}{2}}{r}\,\Deltah^{[r]}
\label{Dh-op}
\,,
\eea
where $\binom{a}{b}$ is the binomial coefficient and, as in Ref.~\cite{Lanata2016},
we adopted the following notation for indicating 
operatorial matrix products:
\bea
[\Dp\bullet\Dh]_{\bR i ab}=
\sum_c \Deltap_{\bR i ac}\Deltah_{\bR i cb}
\eea
and the powers:
\begin{align}
[\Dp]^{\left[l\right]}_{\bR i ab} &=\!\!
\sum_{c_1,..,c_{l-1}}
\Deltap_{Ri ac_1}\Deltap_{\bR i c_1c_2}\,...\,\Deltap_{\bR i c_{l-1}b}\\~
[\Dh]^{\left[l\right]}_{\bR i ab} &=\!\!
\sum_{c_1,..,c_{l-1}}
\Deltah_{Ri ac_1}\Deltah_{\bR i c_1c_2}\,...\,\Deltah_{\bR i c_{l-1}b}\\~
[\hat{\Delta}_p]^{[0]}_{\bR i ab}
&=[\hat{\Delta}_h]^{[0]}_{\bR i ab} = \delta_{ab} \hat{\mathds{1}}
\,.
\end{align}
Note that Eq.~\eqref{R-operatorial} reduces to the expression previously derived in Ref.~\cite{Lanata2016} for $\mathcal{B}=1$.

The fact that Eq.~\eqref{R-operatorial} is equivalent to Eq.~\eqref{quadratic-R}
within the physical subspace is due to the fact that physical states contains only a single Boson, $\forall\,\mathcal{B}$.
In fact, since the Bosonic operators in Eq.~\eqref{R-operatorial} are normally ordered, the matrix elements of Eq.~\eqref{R-operatorial} between states with a single Boson are zero for all terms involving a product of more than 1 creation (or annihilation) operator.
Therfore, neither the terms proportional to $(C_{i,\Gamma\Gamma'}^{\frac{1}{2}}-1)$
nor the terms involving $\Deltap^{[r]}$ or $\Deltah^{[r]}$ with $r>1$ can contribute to matrix elements between physical states.

In summary, we showed that Eq.~\eqref{hubb} can be exactly reformulated in terms of any of the following Hamiltonians:
\begin{align}
\hat{\underline{H}}_{\mathcal{B}} =\sum_{\bR \bRp ij} 
\sum_{\alpha=1}^{{\nu_i}} 
\sum_{\beta=1}^{{\nu_j}} 
{t}_{\bR\bRp,ij}^{\alpha\beta}\,
\underline{c}^\dagger_{\bR i\alpha}
\underline{c}^\dagga_{\bRp j\beta}+
\sum_{\bR,i}
\hat{\underline{H}}_{\bR i}^{\text{loc}}
\,,
\label{hubb-risb}
\end{align}
where the equality is valid $\forall\,\mathcal{B}$ and the physical subspace is identified by the Gutzwiller constraints,
see Eqs.~\eqref{gc1} and \eqref{gc2}.
In the next section we will also show that the resulting g-RISB theory is equivalent to the g-GA at the mean field level.

\section{Mean-field approximation to the $\text{g-RISB}$ theory}\label{MF-section}

Following Ref.~\cite{Lanata2016}, here we derive the mean-field approximation to the g-RISB theory from a variational perspective.

\subsection{The mean-field variational space}\label{mf-varsp}

We consider the most general wavefunction within the Fock space $\underline{\mathcal{F}}$ represented as follows:
\be
\ket{\underline{\Psi}}=\ket{\Psi_0}\otimes \ket{\phi}\,,
\label{mf-ansatz}
\ee
where $\ket{\Psi_0}$ is a generic normalized Fermionic state and $\ket{\phi}$ is a Bosonic coherent state.

For simplicity, here we focus on translationally-invariant solutions, i.e., we assume that both $\ket{\Psi_0}$ and $\ket{\phi}$ are translationally-invariant.
The latter condition can be formalized as follows:
\be
\ket{\phi} \propto e^{\sum_{\bR i}
\sum_{\Gamma n}
[\phi_{i}]_{\Gamma n} \phi^\dagger_{\bR i \Gamma n}
}\,\ket{0}
\label{cohstate}
\ee
where the entries $[\phi_i]_{\Gamma n}$ of the matrices $\phi_i$ (called "slave-Boson amplitudes"), which are the eigenvalues of the Bosonic annihilation operators $\Phi_{\bR i \Gamma n}$, are assumed to be independent of $\bR$.
Note that, within the g-RISB, the matrices $\Phi_{\bR i \Gamma n}$ are not square but rectangular, 
as $\Gamma\in\{0,..,2^{\nu_i}-1\}$, while $n\in\{0,..,2^{\mathcal{B}\nu_i}-1\}$.

Within the mean-field approximation, the Gutzwiller constraints (Eqs.~\eqref{gc1} and \eqref{gc2}) are assumed to be satisfied only in average, i.e., we assume that:
\begin{align}
    \Av{\underline{\Psi}}{\underline{K}^0_{\bR i}}
    &=\Tr[\phi^\dagger_i\phi_i]-1=0
    \label{c1-mf}
    \\
    \Av{\underline{\Psi}}{\underline{K}_{\bR i}^{ab}}
    &=
    \Tr\big[
    \phi^\dagger_i\phi^\dagga_i\,
    \tilde{F}_{ia}^\dagger\tilde{F}_{ib}^\dagga
    \big]
    \!-\!
    \Av{\Psi_0}{\fc_{\bR i a}\fa_{\bR i b}}
    =0
    \label{c2-mf}
    \,,
\end{align}
where $\Tr$ is the trace.
The equations above constrain the variational parameters, i.e., $\ket{\Psi_0}$ and the slave-boson amplitudes $[\phi_i]_{\Gamma n}$.

\subsection{Time-independent mean-field g-RISB theory}

At the mean-field level, the g-RISB approximation to the ground state of $\hat{H}$ corresponds to finding the minimum of the variational energy:
\begin{align}
    \mathcal{E}_{\mathcal{B}}
    =\Av{\underline{\Psi}}{\hat{\underline{H}}_{\mathcal{B}}}
    \,,
\end{align}
see Eq.~\eqref{hubb-risb},
with respect to all mean-field states defined above in Sec.~\ref{mf-varsp}.

From the definition of $\hat{\underline{H}}_{\mathcal{B}}$ in Eq.~\eqref{hubb-risb}, it follows immediately that:
\begin{align}
    \mathcal{E}_{\mathcal{B}}
    &=
    \big\langle{\Psi}_0\big|
    \sum_{\bR\bRp ij}
    \sum_{a=1}^{\mathcal{B}{\nu}_i}
    \sum_{b=1}^{\mathcal{B}{\nu}_j}
    \left[\R_i^\dagga t_{\bR\bRp,ij}\R^{\dagger}_j\right]_{ab}
    \fc_{\bR ia}\fa_{\bR' jb}
    \big|{\Psi}_0\big\rangle
    \nonumber \\
    &+
    N \sum_{\bR i}
    \Tr\big[
    \phi_i^\dagga \phi_i^\dagger\,
    H^{\text{loc}}_i
    \big]
    \,,
    \label{E-mf}
\end{align}
where $N$ is the number of unit cells $\bR$ and we have introduced the $\nu_i\times\nu_j$ matrices $t_{\bR\bRp,ij}$, with entries:
\be
[t_{\bR\bRp,ij}]_{\alpha\beta} = {t}_{\bR\bRp,ij}^{\alpha\beta}
\ee
(see Eq.~\eqref{hubb}), and the $\mathcal{B}\nu_i\times \nu_i$ "renormalization matrices" $\R_i$, with entries:
\be
[\R_i]_{a\alpha}=\Av{\phi}{\hat{\mathcal{R}}_{\bR i a\alpha}}
\label{Ri-intro}
\,,
\ee
where the bosonic operators $\hat{\mathcal{R}}_{\bR i a\alpha}$ are defined in Eq.~\eqref{R-operatorial}.
Note that $\R_i$ does not depend on $\bR$ because of the translational invariance of $\ket{\phi}$, see Eq.~\eqref{cohstate}.

Since the bosonic operators in Eq.~\eqref{R-operatorial} are normally ordered, Eq.~\eqref{Ri-intro} can be readily evaluated by replacing all slave-boson operators with the corresponding coherent-state eigenvalues.
This gives the following expression:
\begin{align}
    [\R_i]_{a\alpha} =
    \sum_{b=1}^{\mathcal{B}\nu_i}\Tr\big[
    \phi_i^\dagger F^\dagger_{i\alpha}\phi_i\tilde{F}_{ib}^\dagga
    \big]
    \big[
    \Delta_i(\mathds{1}-\Delta_i)
    \big]^{-\frac{1}{2}}_{ba}
    \label{gc0-mf}\,,
\end{align}
where:
\begin{align}
    [{\Delta}_i]_{ab}=\langle \Psi_0 | \fc_{\bR ia} \fa_{\bR ib} | \Psi_0 \rangle
    \label{Delta-mf}
\end{align}
and the symbol $\mathds{1}$ in Eq.~\eqref{gc0-mf} indicates the $\mathcal{B}\nu_i\times\mathcal{B}\nu_i$ identity matrix.
In fact, because of Eq.~\eqref{c1-mf}, we have:
\begin{align}
    \Av{\phi}{
    \sum_{\Gamma''=0}^{2^{\nu_i}-1}
\sum_{n''=0}^{2^{\mathcal{B}\nu_i}-1}
\Phi^\dagger_{\bR i \Gamma'' n''}
\Phi^\dagga_{\bR i \Gamma'' n''}
}=1
\,,
\end{align}
and, because of the definition [Eq.~\eqref{Delta-mf}] and Eq.~\eqref{c2-mf}, we also have:
\begin{align}
    \Av{\phi}{\Deltap_{\bR i ab}}&= 
    \Tr\big[
    \phi_i^{\dagger}\phi_i^{\dagga}
    \tilde{F}_{ia}^\dagger\tilde{F}_{ib}^\dagga
    \big]=
    [{\Delta}_i]_{ab}
    \\
    \Av{\phi}{\Deltah_{\bR i ab}}&=\delta_{ab}-[{\Delta}_i]_{ab}
\,.
\end{align}

\subsection{Time-independent mean-field Lagrange function}

Here we show that the problem of calculating the energy minimum  
of Eq.~\eqref{E-mf} subject to the constraints~\eqref{c1-mf} and \eqref{c2-mf}, can be conveniently formulated as a quantum-embedding theory $\forall\, \mathcal{B}\geq 1$.

Following Refs.~\cite{Gmethod,Our-PRX,Lanata2016}, this can be accomplished by introducing the following Lagrange function:
\begin{align}
    \Lag &= 
    \Lag_{\text{qp}} + \sum_i\Lag_{\text{emb}}^{i}
    +\Lag_{\text{mix}}
    \,,
    \label{lagmfstatic}
\end{align}
where:
\begin{align}
    \Lag_{\text{qp}}=
    N^{-1}
    \left(
    \Av{\Psi_0}{\hat{H}_{\text{qp}}-{\mathcal{E}}}
    +{\mathcal{E}}
    \right)
    \,,
    \label{lagmfstaticqp}
\end{align}
and:
\begin{widetext}
\begin{align}
    \hat{H}_{\text{qp}}&=
    \sum_{\bR\bRp ij}
    \sum_{a=1}^{\mathcal{B}{\nu}_i}
    \sum_{b=1}^{\mathcal{B}{\nu}_j}
    \big[\R_{i}^\dagga(t) t_{\bR\bRp,ij}\R^{\dagger}_{ j}(t)\big]_{ab}
    \,
    \fc_{\bR ia} \fa_{\bRp jb}
    +
    \sum_{\bR i} \sum_{ab}
    [{\Lambda}_i]_{ab}\,
    \fc_{\bR ia} \fa_{\bR ib}
    \,,
    \label{hqp}
    \\
    \Lag_{\text{emb}}^{i} &= 
     \Tr\big[\phi_i^{\dagga}\phi_i^{\dagger}\,H^{\text{loc}}_i\big]
    +
    \sum_{a=1}^{\mathcal{B}\nu_i}\sum_{\alpha=1}^{\nu_i}
    \left(
    [\mathcal{D}_i]_{a\alpha}
    \Tr\left[
    \phi_i^{\dagger}
    F^\dagger_{i\alpha}
    \phi_i^{\dagga}
    \tilde{F}_{ia}
    \right]
    +\text{c.c.}
     \right)
    +\sum_{a,b=1}^{\mathcal{B}\nu_i}
     {[\Lambda}^c_i]_{ab}
     \Tr\big[
    \phi_i^{\dagger}\phi_i^{\dagga}
    \tilde{F}_{ia}^\dagger\tilde{F}_{ib}^\dagga
    \big]
    +
    \mathcal{E}^c_i
    \left(
    1- 
    \Tr\big[\phi_i^{\dagger}\phi_i^{\dagga}\big]
    \right)
    \,,
    \label{Lemb-amplitudes}
    \\
    \Lag_{\text{mix}}&=
    -\sum_{i}\left[
    \sum_{a,b=1}^{\mathcal{B}\nu_i}
    \left(
    [\Lambda^{\dagga}_i]_{ab} +[\Lambda^{c}_i]_{ab}
    \right)[\Delta_i]_{ab}
    +
    \sum_{c,a=1}^{\mathcal{B}\nu_i}\sum_{\alpha=1}^{\nu_i}
    \left(
    [\mathcal{D}_i]_{a\alpha}
    [\mathcal{R}_i]_{c\alpha}
    \left[
    \Delta_i
    \left(
    \mathds{1}
    -\Delta_i
    \right)\right]_{ca}^{\frac{1}{2}}
    +\text{c.c.}
    \right)
    \right]
    \,.
\end{align}

\end{widetext}

The Lagrange function above is derived by adding to the energy function [Eq.~\eqref{E-mf}] (divided by $N$) the following terms:
\begin{itemize}

\item the normalization condition $\langle \Psi_0 | \Psi_0 \rangle = 1$ is enforced with the Lagrange multiplier ${\mathcal{E}}$, by introducing:
\begin{align}
N^{-1}\mathcal{E}\left(
1-\langle \Psi_0 | \Psi_0 \rangle
\right)\,;
\end{align}

\item the Gutzwiller constraint [Eq.~\eqref{c1-mf}] is enforced with the Lagrange multipliers ${\mathcal{E}^c_i}$, by introducing:
\begin{align}
\mathcal{E}^c_i
    \left(
    1- 
    \Tr\big[\phi_i^{\dagger}\phi_i^{\dagga}\big]
    \right)
    \,;
\end{align}

\item the entries of the matrices ${\Delta}_i$ (see Eq.~\eqref{Delta-mf}) are promoted to independent variables with the matrices of Lagrange multipliers ${\Lambda}_i$, by introducing:
\begin{align}
    \sum_{a,b=1}^{\mathcal{B}\nu_i}
    \left(
    N^{-1}\sum_{\bR}
    \Av{\Psi_0}{\fc_{\bR ia} \fa_{\bR ib}}
    - [\Delta_i]_{ab}
    \right)
    [\Lambda^{\dagga}_i]_{ab}
    \,;
\end{align}

\item the Gutzwiller constraints [Eq.~\eqref{c2-mf}] are enforced with the matrices of Lagrange multipliers ${\Lambda}^c_i$, by introducing:
\begin{align}
    \sum_{a,b=1}^{\mathcal{B}\nu_i}
    \left(
    \Tr\big[
    \phi_i^{\dagger}\phi_i\,
    \tilde{F}_{ia}^\dagger\tilde{F}_{ib}^\dagga
    \big]
    - [\Delta_i]_{ab}
    \right)
    [\Lambda^{c}_i]_{ab}
    \,;
\end{align}

\item the entries of the matrices $\mathcal{R}_i$ are promoted to independent variables with the matrix of Lagrange multipliers $\mathcal{D}_i$, by introducing:
\begin{align}
    \sum_{a=1}^{\mathcal{B}\nu_i}\sum_{\alpha=1}^{\nu_i}
    [\mathcal{D}_i]_{a\alpha}
    &
    \left(
    \Tr\left[
    \phi_i^{\dagger}
    F^\dagger_{i\alpha}
    \phi_i^{\dagga}
    \tilde{F}_{ia}
    \right]
    \right.
    \nonumber\\
    &\left.
    -\sum_{c=1}^{\mathcal{B}\nu_i}
    [\mathcal{R}_i]_{c\alpha}
    \left[
    \Delta_i
    \left(
    \mathds{1}
    -\Delta_i
    \right)\right]_{ca}^{\frac{1}{2}}
    \right)
    \,.
\end{align}

\end{itemize}

\subsubsection*{Quantum-embedding mapping}

Following Ref.~\cite{Our-PRX}, we introduce the so called "embedding states," which are related to the g-RISB amplitudes as follows:
\begin{align}
    \ket{\Phi_i}&=\sum_{\Gamma=0}^{2^{\nu_i}-1} \,\sum_{n=0}^{2^{\mathcal{B}{\nu}_i}-1}
    e^{i\frac{\pi}{2}N(n)(N(n)-1)}
    \,[\phi_i]_{\Gamma n} 
    \nonumber\\
    &\qquad\qquad\qquad\qquad\quad
    \ket{\Gamma;i}\otimes U_{\text{PH}}\ket{n;i}
    \,,
    \label{ehstates}
\end{align}
where:
\begin{align}
    \ket{\Gamma;i}&= [\hat{c}^\dagger_{i1}]^{q_1(\Gamma)} ...
[\hat{c}^\dagger_{i q_{{\nu}_i}}]^{q_{{\nu}_i}(\Gamma)}\,\ket{0}
    \\
    \ket{n;i}&= [\hat{f}^\dagger_{i1}]^{q_1(n)} ...
[\hat{f}^\dagger_{i q_{\mathcal{B}{\nu}_i}}]^{q_{\mathcal{B}{\nu}_i}(n)}\,\ket{0}
    \,,
\end{align}
$U_{\text{PH}}$ is a particle-hole transformation acting over the $\ket{n;i}$ states and $N(n)$ was defined in Eq.~\eqref{Nofn}.

The set of all embedding states represented in Eq.~\eqref{ehstates} constitute a Fock space, corresponding to an "impurity" (generated by the Fermionic degrees of freedom $\hat{c}_{i\alpha}$, $\alpha\in\{1,..,\nu_i\}$) and a "bath" (generated by the Fermionic degrees of freedom $\hat{f}_{i a}$, $a\in\{1,..,\mathcal{B}{\nu}_i\}$).
Note that the condition [Eq.~\eqref{half-filling-EH}] amounts to assume that the embedding states $\ket{\Phi_i}$ have a total of $(\mathcal{B}{\nu}_i+\nu_i)/2$ electrons, i.e., that they are half-filled ---which is the same condition previously assumed in Refs.~\cite{Ghost-GA, ALM_g-GA}, within the g-GA framework.

It can be readily verified by inspection that, within these definitions, all of the terms of $\Lag^i_{\text{emb}}$ (see Eq.~\eqref{Lemb-amplitudes}) can be replaced with:
\bea
\Lag_{\text{emb}}^{i}=
\Av{\Phi_i}{\h_{i}^{\text{emb}}}
+\mathcal{E}_i\!\left(1-\langle \Phi_i | \Phi_i \rangle
\right)
\label{lagmfstaticemb}
\,,
\eea
where:
\begin{widetext}
\begin{align}
    \h_{i}^{\text{emb}}&= 
    \hat{H}^{\mathrm{loc}}_{i}\big[\hat{c}^{\dagger}_{ i\alpha},\hat{c}^{\phantom{\dagger}}_{i\alpha} \big] 
    +
    \sum_{a=1}^{\mathcal{B}{\nu}_i}\sum_{\alpha=1}^{\nu_i}
    \left(
    \left[\D_i\right]_{a\alpha}
    \hat{c}^{\dagger}_{i\alpha}\hat{f}^{\phantom{\dagger}}_{ia}
    + \text{H.c.}
    \right) 
    +\sum_{a,b=1}^{\mathcal{B}{\nu}_i}
   \left[\Lambda^c_i\right]_{ab}\hat{f}^{\phantom{\dagger}}_{ib}\hat{f}^{\dagger}_{ia}
   \,,
   \label{hemb}
\end{align}
\end{widetext}
where $\hat{H}^{\mathrm{loc}}_{i}$ is obtained from the operator
$\hat{H}_{\bR i}^{\text{loc}}$ appearing in Eq.~\eqref{hubb} by replacing $\ca_{\bR i\alpha}$ with $\hat{c}^{\phantom{\dagger}}_{i\alpha}$ $\forall\,\alpha$.

The mean-field Lagrange function derived above
(Eqs.~\eqref{lagmfstatic}, \eqref{lagmfstaticqp}, \eqref{hqp}, \eqref{lagmfstaticemb} and \eqref{hemb})
coincides with the g-GA Lagrange function, see the supplemental material of Ref.~\cite{ALM_g-GA}.
This proves that, as we claimed in the introduction, the g-RISB reduces to the g-GA at the mean-field level.

\subsection{Time-dependent mean-field action}

For completeness, here we show that from the mean-field g-RISB variational ansatz defined in Sec.~\ref{mf-varsp} it is also possible to define a mean-field theory for the real-time dynamics.
This can be accomplished by extremizing the following Dirac-Frenkel action:
\begin{align}
S &= N^{-1} \int_{t_i}^{t_f} \Av{\underline{\Psi}(t)}{i\partial_t-\hat{\underline{H}}_{\mathcal{B}}}
dt
\,,
\label{td-action}
\end{align}
where $\underline{\Psi}(t)$ is of the form represented in Eq.~\eqref{mf-ansatz}.

Note that, in principle, Eq.~\eqref{td-action} should be extremized only with respect to variations $\ket{\delta \underline{\Psi}(t)}$ (with boundary conditions $\ket{\delta \underline{\Psi}(t_i)}=\ket{\delta \underline{\Psi}(t_f)}=0$) lying within the subset of variational states satisfying the Gutzwiller constraints, see Eqs.~\eqref{c1-mf} and \eqref{c2-mf}.
On the other hand, since:
\be
\Av{\underline{\Psi}(t)}{\hat{\underline{H}}_{\mathcal{B}}}
=\Av{\hat{\mathcal{G}}\underline{\Psi}(t)}{\hat{\underline{H}}_{\mathcal{B}}}
\ee
for every Lie-group gauge transformation $\hat{\mathcal{G}}$ generated by the Gutzwiller-constraint operators [Eqs.~\eqref{gc1} and \eqref{gc2}], the Gutzwiller constraints are preserved automatically (in average) by the dynamics.
In other words, $\Av{\underline{\Psi}(t)}{\underline{K}^0_{\bR i}}$ and
$\Av{\underline{\Psi}(t)}{\underline{K}_{\bR i}^{ab}}$ are the conserved quantities associated with the gauge group of $\hat{\underline{H}}_{\mathcal{B}}$.
Therefore, as in the time-dependent GA~\cite{tdGA-schiro,tdGA-lanata},
it is possible to extremize Eq.~\eqref{td-action} with respect to all variations $\ket{\delta \underline{\Psi}(t)}$, as the Gutzwiller constraints are satisfied automatically.

By expressing explicitly the integrand of Eq.~\eqref{td-action} in terms of the variational parameters, using the identities derived above in Sec.~\ref{MF-section}, we obtain the folowing expression:
\begin{align}
S &=\int_{t_i}^{t_f}
    \left\{
     N^{-1} \Av{\Psi_0(t)}{
    i\partial_t - 
    \hat{H}_{\text{qp}}(t)
    }
    \right.\nonumber\\&\left.
+
    \sum_i\Av{\Phi_i(t)}{i\partial_t\!-\!\hat{H}^{\mathrm{loc}}_{i}}
\right\} 
dt
\,,
\label{action-nonlinear}
\end{align}
where:
\begin{align}
    \hat{H}_{\text{qp}}(t)&=
    \sum_{\bR\bRp ij}
    \sum_{a=1}^{\mathcal{B}{\nu}_i}
    \sum_{b=1}^{\mathcal{B}{\nu}_j}
    \big[\R_{i}^\dagga(t) t_{\bR\bRp,ij}\R^{\dagger}_{ j}(t)\big]_{ab}
    \,
    \fc_{\bR ia} \fa_{\bRp jb}
    \,.
    \label{hqp-td}
\end{align}
Note that the dynamical variables in Eq.~\eqref{action-nonlinear} are only $\ket{\Psi_0(t)}$, the embedding states $\ket{\Phi_i(t)}$ and their respective time derivatives, while the entries of $\mathcal{R}_i(t)$ are expressed in terms of $\ket{\Phi_i(t)}$ as follows:
\begin{align}
    &[\R_i(t)]_{a\alpha} =\Av{\phi(t)}{\hat{\mathcal{R}}_{\bR i a\alpha}}
    \nonumber\\
    &\;\;\;=\sum_{b=1}^{\mathcal{B}\nu_i}\Tr\big[
    \phi_i^\dagger(t) F^\dagger_{i\alpha}\phi_i(t)\tilde{F}_{ib}^\dagga
    \big]
    \big[
    \Delta_i(t)(\mathds{1}-\Delta_i(t))
    \big]^{-\frac{1}{2}}_{ba}
    \nonumber\\
    &\;\;\;=\sum_{b=1}^{\mathcal{B}\nu_i}
    \Av{\Phi_i}{\hat{c}^{\dagger}_{i\alpha}\hat{f}^{\phantom{\dagger}}_{ia}}
    \big[
    \Delta_i(t)(\mathds{1}-\Delta_i(t))
    \big]^{-\frac{1}{2}}_{ba}
    \,,
    \label{gc0-mf-td}
\end{align}
where:
\begin{align}
    [{\Delta}_i(t)]_{ab}&=
    \Tr\big[
    \phi^\dagger_i(t)\phi^\dagga_i(t)\,
    \tilde{F}_{ia}^\dagger\tilde{F}_{ib}^\dagga
    \big]
    \nonumber\\
    &=\Av{\Phi_i(t)}{\hat{f}^{\phantom{\dagger}}_{ib}\hat{f}^{\dagger}_{ia}}
    \label{Delta-mf-td}
    \,.
\end{align}

To derive the time-dependent g-RISB equations of motion, 
it is convenient to reformulate the problem  by promoting to independent dynamical variables the entries of $\Delta_i(t)$ with matrices of Lagrange multipliers $\Lambda^c_i(t)$ and by promoting to independent dynamical variables the entries of $\R_i(t)$ with matrices of Lagrange multipliers $\D_i(t)$ (as we did above for the static theory).
This leads to the following action:
\begin{widetext}
\begin{align}
S =\int_{t_i}^{t_f}
    &\left\{
    N^{-1} \Av{\Psi_0(t)}{
    i\partial_t - 
    \hat{H}_{\text{qp}}(t)
    }
+ \sum_i\Av{\Phi_i(t)}{i\partial_t - \h_{i}^{\text{emb}}(t)}
\right.
\nonumber\\
& \left.
    +\sum_{i}\left[
    \sum_{a,b=1}^{\mathcal{B}\nu_i}
    [\Lambda^{c}_i(t)]_{ab}
    [\Delta_i(t)]_{ab}
    +
    \sum_{c,a=1}^{\mathcal{B}\nu_i}\sum_{\alpha=1}^{\nu_i}
    \left(
    [\mathcal{D}_i(t)]_{a\alpha}
    [\mathcal{R}_i(t)]_{c\alpha}
    \left[
    \Delta_i(t)
    \left(
    \mathds{1}
    -\Delta_i(t)
    \right)\right]_{ca}^{\frac{1}{2}}
    +\text{c.c.}
    \right)
    \right]
\right\} 
dt
\,,
\label{action-linear}
\end{align}
\end{widetext}
where $\h_{i}^{\text{emb}}(t)$ is defined in terms of $\mathcal{D}_i(t)$ and 
$\Lambda^c_i(t)$ as in Eq.~\eqref{hemb}.

We point out that Eq.~\eqref{action-linear} is equivalent to Eq.~\eqref{action-nonlinear}, but the dynamical variables are not only $\ket{\Psi_0(t)}$, $\ket{\Phi_i(t)}$ and their time derivatives, but also
${\Delta}_i(t)$, $\Lambda^c_i(t)$, $\R_i(t)$ and $\D_i(t)$.
The advantage of Eq.~\eqref{action-linear} is that the second term
represents the Dirac-Frenkel action for a time-dependent embedding Hamiltonian,
which depends explicitly on $\ket{\Phi_i(t)}$ only linearly.
Therefore, the time-dependent g-RISB dynamics is described in terms of 2 coupled time-dependent Schr\"odinger equations (one for 
$\ket{\Psi_0(t)}$ and one for $\ket{\Phi_i(t)}$).

Note that Eq.~\eqref{action-linear} reduces to the time-dependent GA action for $\mathcal{B}=1$, see Refs.~\cite{tdGA-schiro,tdGA-lanata}, while the number of bath sites in the embedding Hamiltonian is $\mathcal{B}>1$ in the g-RISB generalization.
Since increasing the value of $\mathcal{B}$ corresponds to extending the variational freedom (both within the g-RISB and within the g-GA frameworks), it
shall be expected to improve systematically the accuracy, as we hope to verify numerically in future work.

\section{Conclusions}

We derived a ground-state and time-dependent theory of multi-orbital electronic systems interacting locally on a lattice (the g-RISB), that reduces to the g-GA at the mean-field level.
This provides an alternative perspective on the g-GA theory, which may pave the way for developing new generalizations.
In particular, since the g-RISB is based on an exact reformulation of the many-body problem, it may lead to practical implementations for calculating systematic corrections ---e.g., using cluster methods~\cite{CDMFT-Jarrell,CDMFT-Potthoff,CDMFT-Lichtenstein}, or generalizing
the path-integral approach for including quantum fluctuations previously developed in Refs.~\cite{SB-fluct-1,SB-fluct-2,SB-fluct-3}.
Since the mean-field g-RISB theory describes the electronic structure in terms of emergent Bloch excitations, and such description
proved to have accuracy comparable with DMFT~\cite{Ghost-GA,ALM_g-GA},
we argue that taking into account perturbatively the residual effective interactions between the g-GA generalized Bloch excitations within the g-RISB framework is a promising route for performing high-precision calculations of strongly-correlated materials beyond the DMFT level, in arbitrary dimension~\cite{nonlocal-corr-review}.

\section*{Acknowledgements}
I thank Tsung-Han Lee for useful discussions.
I gratefully acknowledge funding from the Novo Nordisk Foundation through the Exploratory Interdisciplinary Synergy Programme project NNF19OC0057790.
I thank support from the VILLUM FONDEN through the Villum Experiment project 00028019 and the Centre of Excellence for Dirac Materials (Grant. No. 11744).


%

\end{document}